\documentclass{article}

\usepackage[preprint]{neurips_data_2023}

\usepackage[utf8]{inputenc} 
\usepackage[T1]{fontenc}    
\usepackage{hyperref}       
\usepackage{url}            
\usepackage{booktabs}       
\usepackage{amsfonts}       
\usepackage{nicefrac}       
\usepackage{microtype}      
\usepackage{xcolor}         
\usepackage{xurl}

\usepackage[utf8]{inputenc} 
\usepackage[T1]{fontenc}    
\usepackage{hyperref}       
\usepackage{url}            
\usepackage{booktabs}       
\usepackage{amsfonts}       
\usepackage{nicefrac}       
\usepackage{microtype}      
\usepackage{xcolor}         
\usepackage{amsmath}
\usepackage{caption}
\usepackage{bbm}
\usepackage{xspace}
\usepackage{multirow}
\usepackage{longtable}
\usepackage{makecell}
\usepackage{listings}
\usepackage{soul}
\usepackage{tikz}
\usetikzlibrary{calc}
\usetikzlibrary{tikzmark}
\usepackage{subcaption}
\usepackage{tabularx}
\usepackage{upquote}
\usepackage{lineno}

\usepackage[capitalize,noabbrev]{cleveref}

\newcommand{\codeID}[1]{{\small \ttfamily #1}}
\newcommand{\dataset}{CodeQueries\xspace}

\lstalias{no_lang}{}

\newcommand{\mytikzmark}[1]{%
  \tikz[overlay,remember picture,baseline] \coordinate (#1) at (0,0) {};}

\newcommand{\highlight}[2]{%
  \draw[red,line width=8pt,opacity=0.3]%
    ([yshift=1pt]#1) -- ([yshift=1pt]#2);%
}

\newcommand{\highlightSF}[2]{%
  \draw[green,line width=8pt,opacity=0.3]%
    ([yshift=1pt]#1) -- ([yshift=1pt]#2);%
}

\def\verythinspace{%
  \ifmmode
    \mskip0.5\thinmuskip
  \else
    \ifhmode
      \kern0.08334em
    \fi
  \fi
}

\newcolumntype{R}[1]{>{\raggedleft\let\newline\\\arraybackslash\hspace{0pt}}m{#1}}
\newcolumntype{L}[1]{>{\raggedright\let\newline\\\arraybackslash\hspace{0pt}}m{#1}}

\newcommand{\unusedimport}{Unused import}
\newcommand{\missinginitduringinit}{Missing call to \codeID{\_\_init\_\_} during object initialization}
\newcommand{\useofnonereturn}{Use of the return value of a procedure}
\newcommand{\eqnotoverriden}{\codeID{\_\_eq\_\_} not overridden when adding attributes}
\newcommand{\wrongargincall}{Wrong number of arguments in a call}
\newcommand{\comparisonusingis}{Comparison using is when operands support \codeID{\_\_eq\_\_}}
\newcommand{\signmismatch}{Signature mismatch in overriding method}
\newcommand{\noncallablecalled}{Non-callable called}
\newcommand{\initcalloverriden}{\codeID{\_\_init\_\_} method calls overridden method}
\newcommand{\conflictingattr}{Conflicting attributes in base classes}
\newcommand{\iterreturnnoniter}{\codeID{\_\_iter\_\_} method returns a non-iterator}
\newcommand{\flaskdebug}{Flask app is run in debug mode}
\newcommand{\unequalhash}{Inconsistent equality and hashing}
\newcommand{\wrongarginclassinst}{Wrong number of arguments in a class instantiation}
\newcommand{\incompleteordering}{Incomplete ordering}

\newcommand{\unusedlocalvar}{Unused local variable}
\newcommand{\baseexception}{Except block handles \codeID{BaseException}}
\newcommand{\impreciseassert}{Imprecise assert}
\newcommand{\vardefmultiple}{Variable defined multiple times}
\newcommand{\testeqnone}{Testing equality to None}
\newcommand{\unreachablecode}{Unreachable code}
\newcommand{\firstparaself}{First parameter of a method is not named \codeID{self}}
\newcommand{\unnecessarypass}{Unnecessary pass}
\newcommand{\doubletypeimport}{Module is imported with \codeID{import} and \codeID{import from}}
\newcommand{\doubleimport}{Module is imported more than once}
\newcommand{\constcompare}{Comparison of constants}
\newcommand{\implicitstringconcat}{Implicit string concatenation in a list}
\newcommand{\unusedloopvar}{Suspicious unused loop iteration variable}
\newcommand{\duplicatedictkey}{Duplicate key in dict literal}
\newcommand{\unnecessaryelse}{Unnecessary \codeID{else} clause in loop}
\newcommand{\firstargsuper}{First argument to super() is not enclosing class}
\newcommand{\redundantassign}{Redundant assignment}
\newcommand{\assertsideeffect}{An assert statement has a side-effect}
\newcommand{\samevarinnestedloop}{Nested loops with same variable}
\newcommand{\importdeprecated}{Import of deprecated module}
\newcommand{\notimplemented}{NotImplemented is not an Exception}
\newcommand{\redundantcomp}{Redundant comparison}
\newcommand{\deprecatedslice}{Deprecated slice method}
\newcommand{\constincondition}{Constant in conditional expression or statement}
\newcommand{\compidentical}{Comparison of identical values}
\newcommand{\importstar}{\codeID{import }$\star$ may pollute namespace}
\newcommand{\unnecessarydelete}{Unnecessary delete statement in function}
\newcommand{\illegalraise}{Illegal raise}
\newcommand{\insecuretemp}{Insecure temporary file}
\newcommand{\moddefaultpara}{Modification of parameter with default}
\newcommand{\usewith}{Should use a \codeID{with} statement}
\newcommand{\specialmethodsign}{Special method has incorrect signature}
\newcommand{\nonstandardexc}{Non-standard exception raised in special method}
\newcommand{\useglobalmodule}{Use of \codeID{global} at module level}
\newcommand{\moddictlocals}{Modification of dictionary returned by locals()}
\newcommand{\incompleteurlsanity}{Incomplete URL substring sanitization}
\newcommand{\unguardednext}{Unguarded next in generator}

\renewcommand{\paragraph}[1]{\textbf{#1}}

\title{CodeQueries: A Dataset of Semantic Queries over Code}

\author{Surya Prakash Sahu\\
Indian Institute of Science\\
\And
Madhurima Mandal\thanks{Now at Myntra.}\\
Indian Institute of Science\\
\And
Shikhar Bharadwaj\thanks{Now at Google Research.}\\
Indian Institute of Science\\
\And
Aditya Kanade\\
Microsoft Research\\
\And
Petros Maniatis\\
Google DeepMind\\
\And
Shirish Shevade\\
Indian Institute of Science\\
}

\begin{document}

\maketitle

\setcounter{footnote}{0}

\begin{abstract}
Developers often have questions about semantic aspects of code they are working on, e.g., ``Is there a class whose parent classes declare a conflicting attribute?''. Answering them requires understanding code semantics such as attributes and inheritance relation of classes. An answer to such a question should identify code spans constituting the answer (e.g., the declaration of the subclass) as well as supporting facts (e.g., the definitions of the conflicting attributes). The existing work on question-answering over code has considered yes/no questions or method-level context. We contribute a labeled dataset, called \dataset, of semantic queries over Python code. Compared to the existing datasets, in \dataset, the queries are about code semantics, the context is file level and the answers are code spans. We curate the dataset based on queries supported by a widely-used static analysis tool, CodeQL, and include both positive and negative examples, and queries requiring single-hop and multi-hop reasoning.

To assess the value of our dataset, we evaluate baseline neural approaches. We study a large language model (GPT3.5-Turbo) in zero-shot and few-shot settings on a subset of \dataset. We also evaluate a BERT style model (CuBERT) with fine-tuning. 
We find that these models achieve limited success on \dataset. \dataset is thus a challenging dataset to test the ability of neural models, to understand code semantics, in the extractive question-answering setting.
\end{abstract}

\section{Introduction}

Extractive question-answering in natural-language settings is a venerable domain of NLP, requiring detailed reasoning about a single reasoning step (``single hop''~\citep{DBLP:conf/emnlp/RajpurkarZLL16})
or multiple reasoning steps (``multi-hop''~\citep{DBLP:conf/emnlp/Yang0ZBCSM18}).
In the context of programming languages, neural question answering over code has not grown to similar complexity:
tasks are either binary yes/no questions~\citep{DBLP:conf/acl/HuangTSG0J0D20} or
range over a localized context (e.g., a source-code method)~\citep{DBLP:conf/wcre/BansalEWM21, DBLP:conf/emnlp/Liu021}. 

Recent results show promise towards neural program analyses around complex concepts
such as program invariants~\citep{DBLP:conf/nips/SiDRNS18, daikonml}, inter-procedural properties~\citep{DBLP:conf/icml/CumminsFBHOL21},
and even evidence of deeper semantic meaning~\citep{semanticmeaning}. 
However, there do not exist semantically rich question-answering datasets requiring reasoning over code, especially for questions with large scope (entire files)
and high complexity (e.g., multi-hop reasoning).
Also, given the criticality of program analysis, it is pertinent to judge neural approaches not only on the answer to a question,
but also on the reasoning or evidence for that answer.

\begin{table}[t]
\begin{minipage}[b]{0.4\columnwidth}
\begin{lstlisting}[basicstyle=\small]
class TCPServer:
    |\mytikzmark{b1Start}|def __init__(self, service, ...): ...

    |\mytikzmark{b2Start}|def serve(self, address): ...

    # Supporting Fact 1
    |\mytikzmark{b3Start}||\mytikzmark{sf1Start}|def acceptConnection(self, conn)|\mytikzmark{sf1End}|: ...
    |\phantom{def acceptConnection(self, conn):}| |\mytikzmark{b3End}|
    |\mytikzmark{b4Start}|def handleConnection(self, conn): ...

class ThreadingMixin:
    # Supporting Fact 2
    |\mytikzmark{b5Start}||\mytikzmark{sf2Start}|def acceptConnection(self, conn)|\mytikzmark{sf2End}|: ...
    |\phantom{t = Thread(target=self.handleConnection, args=(conn,))}| |\mytikzmark{b5End}|
# Answer Span
|\mytikzmark{spanStart1}||\mytikzmark{b6Start}|class ThreadedTCPServer|\mytikzmark{midpoint}|(|\mytikzmark{spanEnd1}|
    |\mytikzmark{spanStart2}|ThreadingMixin, TCPServer)|\mytikzmark{spanEnd2}|:
    pass
\end{lstlisting}

\begin{tikzpicture}[remember picture, overlay, x=1mm, y=1mm]
\highlight{spanStart1}{spanEnd1};
\highlight{spanStart2}{spanEnd2};
\highlightSF{sf1Start}{sf1End};
\highlightSF{sf2Start}{sf2End};
\end{tikzpicture}
\captionof{figure}{Example code labeled with the answer and supporting-fact spans for the conflicting-attributes query.}
\label{fig:conflicting-attributes-compressed}
\end{minipage}
\hfill
\begin{minipage}[b]{0.55\columnwidth}
\includegraphics[width=\columnwidth]{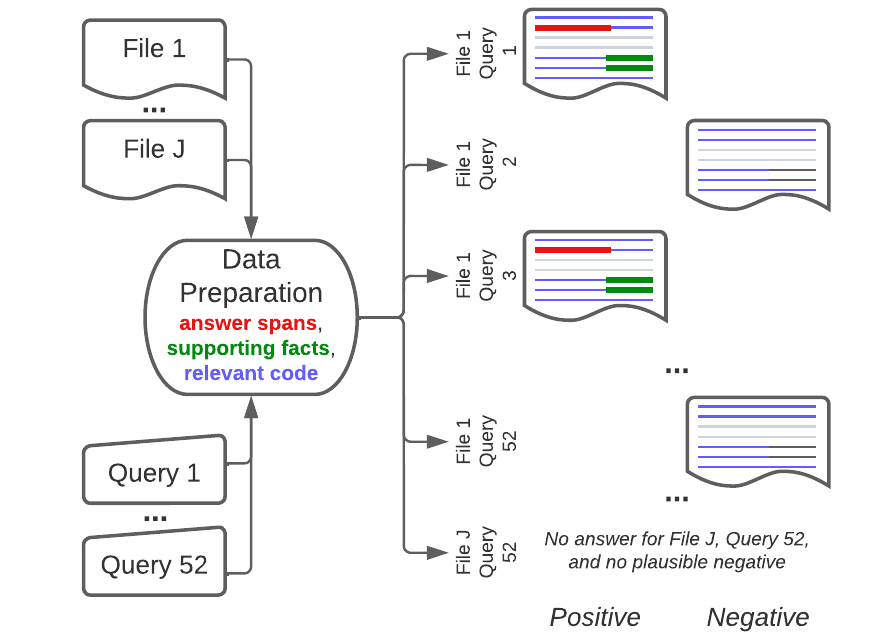}
\captionof{figure}{Methodology for preparing the \dataset dataset.
All source-code files are analyzed against each of the 52 CodeQL queries to gather multiple positive and negative examples for that query.
We derive answer spans, supporting-fact spans and code relevant for answering the query for each example. The details are discussed in Section~\ref{sec:dataset-preparation}.}
\label{fig:data-preparation-setup}
\end{minipage}
\vspace*{-10pt}
\end{table}

In this work, we set out to build a labeled dataset, called \emph{\dataset}, for extractive question-answering over code.
The queries are described in English and the context is provided by the contents of a source-code file. If a file does not contain code spans matching the queried pattern then the answer spans is an empty set. These are \emph{negative examples}. \emph{Positive examples} provide \emph{answer spans} in the file. Some queries require reasoning about multiple facts. For them, the \emph{supporting facts} are also identified as code spans in the file. As an example, consider a query about existence of ``conflicting attributes in base classes''.
Figure~\ref{fig:conflicting-attributes-compressed} shows a positive example labeled with answer and supporting-fact spans. 
The subclass \codeID{ThreadedTCPServer} inherits from the two base classes \codeID{ThreadingMixin} and \codeID{TCPServer}, both of which define method \codeID{acceptConnection}. Since both superclasses define the same method, there is a conflict in resolving method \codeID{acceptConnection} invoked on instances of \codeID{ThreadedTCPServer}.
As shown in the figure, the declaration of the subclass constitutes the answer span and the declarations of the conflicting attribute in the superclasses constitute supporting facts.

There are two difficulties in constructing such a dataset: 1)~identifying semantic queries that are representative of developers' requirements and 2)~deriving labels. We overcome these difficulties by basing our dataset creation on queries supported by a widely-used static analysis tool, CodeQL\footnote{\url{https://codeql.github.com/}}~\citep{DBLP:conf/ecoop/AvgustinovMJS16}. We identify 52 public CodeQL queries that produce highest number of answers on files in a common corpus of Python code~\citep{raychev2016probabilistic}. Each CodeQL query identifies a semantic aspect of code related to correctness, reliability, maintainability or security of code through program analysis. Among the 52 queries, 15 require \emph{multi-hop reasoning} and 37 require \emph{single-hop reasoning}. For instance, the example in Figure~\ref{fig:conflicting-attributes-compressed} requires multi-hop reasoning across three classes.

Each CodeQL query is evaluated by the CodeQL engine on a relational representation of code (similar to how a database query is evaluated by a database engine). We extract answer and supporting-fact spans from the analysis results. 
Since there can be multiple files in the corpus with code that matches a query, we can gather multiple positive examples per query; e.g., several instances of conflicting attributes from different source-code files. We also include code on which the queries do not return any answer spans (negative examples) so that a model can learn to predict when the code does not have the queried pattern (e.g., absence of a buggy code pattern). These are analogous to the no-answer~\citep{clark2017simple} or unanswerable scenarios~\citep{DBLP:conf/acl/RajpurkarJL18}.
The English descriptions of the CodeQL queries, provided in the CodeQL documentation, are used in the natural-language queries in our dataset.
For example, the ``conflicting attributes in base classes'' query\footnote{\url{https://codeql.github.com/codeql-query-help/python/py-conflicting-attributes/}}  is of the form ``When a class subclasses multiple base classes, attribute lookup is performed from left to right amongst the base classes. ... this means that if more than one base class defines the same attribute ... may not be the desired behavior ...''.
Thus, a neural model will be required to analyze code semantics from the analysis intent described in natural language.
Figure~\ref{fig:data-preparation-setup} shows the data preparation setup. 
\dataset contains 34,662 positive examples and 52,613 negative examples.

To assess the value of our dataset, we consider various baseline neural approaches, varying in architectural choices, evaluation methods and the presence of supporting facts.
Specifically, we study the ability of a large language model (GPT3.5-Turbo), that has seen extensive natural language and code, to answer semantic queries
with various amounts of prompting on a subset of \dataset.
We also study a much smaller but more custom model, fine-tuned from CuBERT~\citep{DBLP:conf/icml/KanadeMBS20}.

We find that these models achieve limited success on \dataset. With zero-shot prompting, GPT3.5-Turbo achieves exact match with the ground-truth answer spans (within pass@10) on 20.84\% of positive examples and detects that 26.77\% negative examples do not contain answer spans. The model performance increases to 32.66\% and 70.08\% respectively when prompted with few-shot examples. The CuBERT model when fine-tuned with limited data achieves exact match on only 3.74\% positive examples. \dataset is thus a challenging dataset that can be used for evaluating current and future neural approaches, on their ability to understand code semantics, in the extractive question-answering setting. 
It can further help understand opportunities to improve model performance.
We have released our code, data and model checkpoints to facilitate future work on the proposed problem of answering
semantic queries over code at \url{https://github.com/thepurpleowl/codequeries-benchmark}.

\section{Related Work}
\paragraph{Natural-language questions and queries about code.}
CoSQA~\citep{DBLP:conf/acl/HuangTSG0J0D20} includes yes/no questions to determine whether a web search query and a method match. \citet{DBLP:conf/wcre/BansalEWM21} and CodeQA~\citep{DBLP:conf/emnlp/Liu021} are two recent works on question-answering over code.
Both consider a method as the code context, and programmatically extract question-answer pairs specific to the method from the method body and comments.
\citet{DBLP:conf/wcre/BansalEWM21} generate questions about method signatures (e.g., what are parameter types),
(mis)matches between a function and a docstring, and function summaries.
CodeQA is generated from code comments using  rule-based templates.
The answers are natural-language sentences extracted from code comments.
The context in our case is larger, file-level; queries are about semantic aspects of code and
may require long chains of reasoning; and answers are spans over code. CS1QA~\citep{lee2022cs1qa} is a dataset of question-answering in an introductory programming course and proposes classification of the question into pre-defined types, identification of relevant source-code lines and retrieval of related QAs.
In an orthogonal direction, natural language queries have been used for code retrieval~\citep{DBLP:conf/icse/GuZ018,Yao_2018,DBLP:journals/corr/abs-1909-09436,DBLP:conf/sigsoft/CambroneroLKS019,DBLP:journals/corr/abs-2008-12193,gu2021cradle}. 

\paragraph{Learning-based program analysis.}
Use of program analysis helps improve software quality. However, implementing analysis algorithms requires expertise and efforts. There is increasing interest in using machine learning for program analysis. Recent work in this direction includes learning program invariants~\citep{DBLP:conf/nips/SiDRNS18,daikonml}, rules for static analysis~\citep{DBLP:conf/cav/BielikRV17}, intra- and inter-procedural data flow analysis~\citep{DBLP:conf/icml/CumminsFBHOL21}, specification inference~\citep{10.1145/3296979.3192383,DBLP:conf/pldi/ChibotaruBRV19}, reverse engineering~\citep{nero}, and type inference~\citep{10.1145/3236024.3236051,DBLP:journals/corr/abs-2004-00348,DBLP:conf/sigsoft/PradelGL020,DBLP:conf/iclr/WeiGDD20,DBLP:journals/corr/abs-2101-04470,peng2022static}. These techniques target specific analysis problems, use specialized program representations or customize learning methods. 
Our work targets semantic queries over code and
presents a uniform extractive question-answering setup for them, wherein the developer intent is expressed in natural language.
Our queries cover diverse program analyses involving forms of type checking, control-flow and data-flow analyses, and many other checks (see the supplementary material for the list of queries). 
\citet{codetrek,codetrek-walks} advocate the use of relational representations of code, as used in CodeQL, in neural modeling and use them on classification tasks.
GitHub has recently launched an experimental service\footnote{\scalebox{0.9}{\url{https://github.blog/2022-02-17-code-scanning-finds-vulnerabilities-using-machine-learning/}}} that uses feature-based machine learning to classify JavaScript and TypeScript code with regards to four common vulnerabilities.

\paragraph{Question-answering over text.}
%
Various datasets for extractive question-answering over text requiring single-hop~\citep{DBLP:conf/emnlp/RajpurkarZLL16} and multi-hop~\citep{DBLP:conf/emnlp/Yang0ZBCSM18} reasoning have been proposed. Our dataset consists of queries requiring single- and multi-hop reasoning over code. Along the lines of prior work~\citep{clark2017simple,DBLP:conf/acl/RajpurkarJL18}, we include negative examples in which the queries cannot be answered with the given context, though the context contains plausible answers~\citep{DBLP:conf/emnlp/Yang0ZBCSM18}.
For improving explainability, we also include in our dataset and models prediction of supporting facts~\citep{DBLP:conf/emnlp/Yang0ZBCSM18}.
We experiment on file-level code which may contain parts that are not relevant to the query. This is analogous to distractor paragraphs~\citep{DBLP:conf/emnlp/Yang0ZBCSM18} and requires the models to deal with spurious information.

\section{Dataset Preparation}
\label{sec:dataset-preparation}

In this section, we describe our methodology for dataset preparation. An example in our dataset is a tuple $(Q, C, A, SF)$ where $Q$ is a query, $C$ is the contents of a Python file, $A$ is the set of answer spans (i.e., code fragments of $C$ that constitute the answer) and $SF$ is the set of supporting-fact spans.

\paragraph{Single-hop and multi-hop queries.}
We evaluated the queries (formalized in the CodeQL query language) from a standard suite of CodeQL~\citep{querySuite} on the redistributable subset~\citep{DBLP:conf/icml/KanadeMBS20} of the ETH Py150 dataset of Python code~\citep{raychev2016probabilistic} (the ETH Py150 Open dataset). These queries are written by experts and identify coding issues pertaining to correctness, reliability, maintainability or security of code.
%
We evaluated each query on individual Python files  (Figure~\ref{fig:data-preparation-setup}). To get a reasonable number of positive examples for each query, we selected queries with at least 50 answer spans in the training split of the ETH Py150 Open dataset. 
We inspected the definition of a query to check whether answering it requires a single reasoning step or multiple reasoning steps, and classified the query accordingly as a \emph{single-hop} or \emph{multi-hop} query.
Out of the 52 queries, 15 are multi-hop and 37 are single-hop. We call these \emph{positive queries}. Note that the formal CodeQL queries are used only for preparing the dataset. We use the English description of a query as the corresponding natural-language query in our dataset.

\paragraph{Positive and negative examples.}
By evaluating a positive query, we identify files containing code spans that satisfy the query definition. These are positive examples for the query. 
Naively, any code on which a query does not return an answer could be viewed as a negative example; for instance, in the case of conflicting attributes (Figure~\ref{fig:conflicting-attributes-compressed}), it would be trivial to answer that there are no conflicting attributes if the code does not contain classes. In natural-language question answering, \citet{DBLP:conf/emnlp/Yang0ZBCSM18} recommend that unanswerable contexts should contain \emph{plausible, but not actual, answers};
otherwise, it is simple to distinguish between answerable and unanswerable contexts~\citep{DBLP:conf/conll/WeissenbornWS17}. Therefore, to obtain negative examples \emph{with plausible answers}, we manually derive logical negations of the CodeQL queries. 
We ensure that a \emph{negative query} identifies code similar to the original (positive) query but which does not satisfy the key properties required for producing an answer to the original query.
For example, the negated version of the conflicting-attributes query finds code containing a class with multiple inheritance (similar to Figure~\ref{fig:conflicting-attributes-compressed}) such that the base classes do \emph{not} have conflicting attributes.
Suppose \codeID{hasMultipleInheritance(c,p1,p2)} and \codeID{haveConflict(p1,p2)} respectively identify a subclass \codeID{c} with two parent classes \codeID{p1} and \codeID{p2}, and check if they have conflicting attributes. The positive query will be \codeID{hasMultipleInheritance(c,p1,p2) and haveConflict(p1,p2)}, whereas the negative query will be \codeID{hasMultipleInheritance(c,p1,p2) and \emph{not} haveConflict(p1,p2)}.
Using results of the negative queries, we derive negative examples.
While the positive queries are already available publicly, we are releasing the negative queries.

\paragraph{Answer and supporting-fact spans.}
We identify the answer and supporting-fact spans from the results produced by the CodeQL engine for each of the positive queries. These spans are of a variety of syntactic patterns, making it non-trivial for a model to identify the right candidates for answering the queries. In all, there are 42 different syntactic patterns of spans such as class declarations, \codeID{with} statements, and list comprehensions. We give the statistics of syntactic patterns of spans in the supplementary material. Note that negative examples do not have answer or supporting-fact spans.

\begin{table}[t]
\begin{minipage}[b]{.55\columnwidth}
\centering
\begin{tabular}{ll|r|r|r}
\hline
\multicolumn{2}{l|}{}       & Train                    & Validation              & Test                     \\ \hline
\multicolumn{1}{l|}{\multirow{2}{*}{Positive}} &
  Min & 34 & 2 & 14 \\ 
\multicolumn{1}{l|}{} & Max &  11,490 &  1,249 &  6,439  \\ 
\multicolumn{1}{l|}{} & \textbf{Total} &  \textbf{20,783}  &  \textbf{2,319} &  \textbf{11,560}  \\\hline
\multicolumn{1}{l|}{\multirow{2}{*}{Negative}} & Min & 29 & 1 & 17 \\ 
\multicolumn{1}{l|}{} & Max &  17,592 &  1,893 &  9,892  \\ 
\multicolumn{1}{l|}{} & \textbf{Total} &  \textbf{31,676} &  \textbf{3,464} &  \textbf{17,473} \\\hline
\end{tabular}
\caption{Dataset statistics.}
\label{tab:file-data-stat}
\end{minipage}
\hfill
\begin{minipage}[b]{0.42\columnwidth}
\includegraphics[width=\columnwidth,height=33mm]{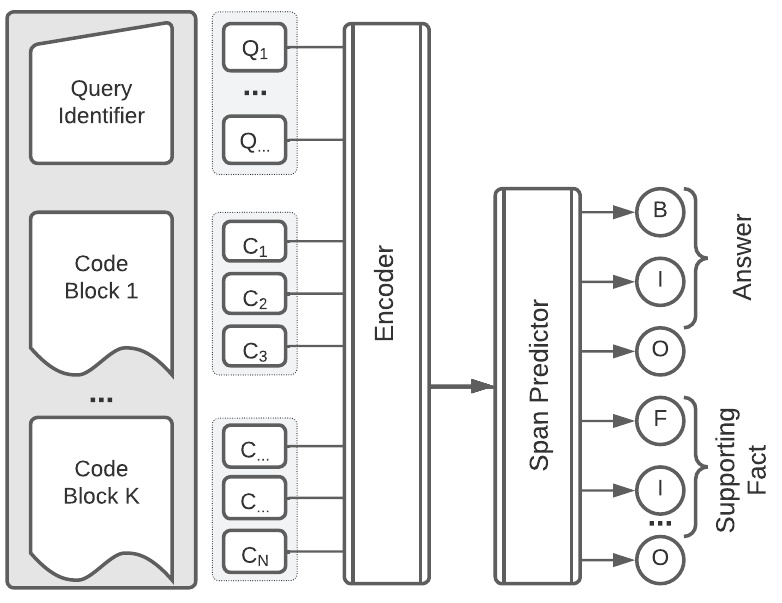}
\vspace*{-10pt}
\captionof{figure}{The span prediction setup.}
\label{fig:model-architecture}
\end{minipage}
\end{table}

\paragraph{Dataset statistics.}
Table~\ref{tab:file-data-stat} gives the dataset statistics according to the splits of the ETH Py150 Open dataset. We place an example derived from a Python file in the same split as the file. 
The Min/Max entries give the number of minimum/maximum examples over individual queries, whereas Total is the sum of examples across all queries. We observed that the query to identify ``unused imports'' produced maximum examples. We provide query-wise statistics in the supplementary material.

\paragraph{Relevant code blocks.}
A CodeQL query produces answers based only on specific parts of code within a file, e.g., a set of classes within the file or a set of methods within a class in the file. We inspect the query definitions and automate extraction of the query-relevant parts from a file. 
Given the query results,
we programmatically obtain the code blocks needed for arriving at the same results for the query. 
We call them \emph{relevant code blocks}. A code block is either a method, all class-level statements (such as attribute definitions) within a class or module-level statements that do not belong to any class or method.
In Section~\ref{subsec:twostep-model}, we describe how this information is used to help the CuBERT model scale to large files by filtering out irrelevant code blocks using a classifier.


\section{Experiment Design}

\dataset is intended as a dataset to analyze semantic understanding of neural models through extractive question-answering over code.
In this work, we evaluate a large language model (LLM) with prompting and a contextual embedding model with fine-tuning, to assess the difficulty level of our dataset.
A full-scale benchmarking of the existing models is \emph{not} an objective of this work.

\subsection{Prompting a Large Language Model}
\label{sec:prompting}

Large language models (e.g.,~\citep{chen_evaluating_2021_Codex,ouyang2022training,li2023starcoder,touvron2023llama,nijkamp2023codegen2,anil2023palm} and others) have shown impressive ability on coding tasks and are capable of zero-shot and few-shot inference~\citep{brown2020language}. We use the GPT3.5-Turbo model~\citep{ouyang2022training} from OpenAI in different settings described below. The complete prompt templates are provided in the supplementary material.

\paragraph{Zero-shot prompting.}
In this setting, we provide the name of the CodeQL query and its English description, both taken from the CodeQL documentation, to the model and instruct it to output answer spans for given code. We require the model to output ``N/A'' if it judges that the code does not have an answer. The contents of a file are provided as the code to be analyzed. The prompt template has the following structure: \texttt{\{Instructions\} \{Code\}}.

\paragraph{Few-shot prompting with BM25 retrieval.}
We provide the same instructions to the model as in the zero-shot prompting but in addition, include a positive and a negative labeled example in the prompt. 
For a query $Q$, we retrieve labeled examples for $Q$ from the training split that are similar to the code to be analyzed, using the BM25 method~\citep{robertson2009probabilistic}. 
The prompt template has the following structure:
\texttt{\{Instructions\} \{Positive example\} \{Negative example\} \{Code\}}. Similar to the zero-shot setting, we require the model to output the answer spans or ``N/A''.
To ensure that we do not overflow the prompt, we minimize the examples by keeping only code blocks that are relevant to the query (see Section~\ref{sec:dataset-preparation}, relevant code blocks).
This optimization is used in the next setting as well.

\paragraph{Few-shot prompting with supporting facts.}
As discussed in Section~\ref{sec:dataset-preparation}, we extract supporting facts from the CodeQL results. In this setting, we evaluate the ability of the LLM to produce both answer and supporting-fact spans. Only positive examples have answer and supporting facts, and therefore this setting is applicable only to the positive examples. The answers to some queries can be determined through local reasoning and they do not have additional supporting facts. 
Our prompt provides instructions to produce answer and supporting facts, and an example with answer and supporting-fact spans. For examples without supporting facts, we mark supporting facts as ``N/A''. The prompt template is:
\texttt{\{Instructions\} \{Example with answer and supporting-fact spans\} \{Code\}}.

\subsection{Fine-tuning a Contextual Embedding Model}
\label{subsec:twostep-model}

\begin{figure}
    \centering
    \includegraphics[width=0.8\columnwidth]{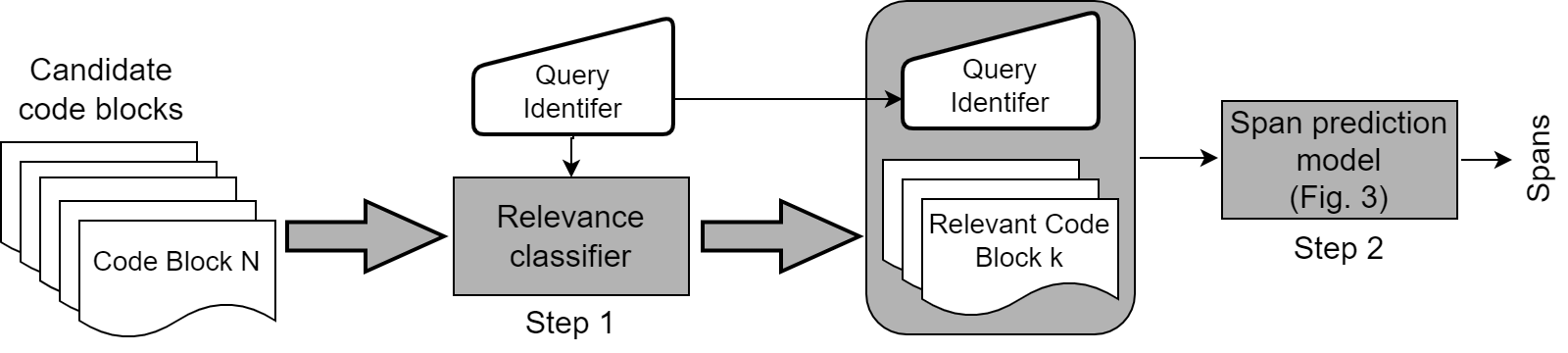}
    \vspace{-3mm}
    \caption{Two-step procedure to handle large-size code containing possibly irrelevant code blocks.}
    \label{fig:two-step}
\end{figure}

\paragraph{Span prediction problem.}
We reformulate the extractive question-answering problem as a problem of classifying code tokens. 
Let $\{B, I, O\}$ respectively indicate \textbf{B}egin, \textbf{I}nside and \textbf{O}utside labels~\citep{DBLP:conf/acl-vlc/RamshawM95}. 
An answer span is represented by a sequence of labels such that the first token of the answer span is labeled by a $B$ and all the other tokens in the span are labeled by $I$'s.
We use an analogous encoding for supporting-fact spans, but we use the $F$ label instead of $B$ to distinguish facts from answers.
Any token that does not belong to a span is labeled by an $O$.
We thus represent multiple answer or supporting-fact spans by a single sequence over $\{B, I, O, F\}$ labels. We call this the \emph{span prediction problem}.
Note that this does not allow overlap between spans, which we have empirically found not to be a problem in our dataset.

\paragraph{Span prediction model.}
We can fine-tune the BERT-style, encoder-based contextual models (e.g.,~\citep{DBLP:conf/icml/KanadeMBS20,DBLP:conf/emnlp/FengGTDFGS0LJZ20,guo2020graphcodebert}) to solve the span prediction problem. We use the CuBERT model~\citep{DBLP:conf/icml/KanadeMBS20} which supports context size of 1K tokens in this work. Figure~\ref{fig:model-architecture} shows the span prediction setup. The input to the model is the unique name of a query (marked as query identifier in the figure) and the code. The whole sequence is preceded with the \codeID{[CLS]} token, similar to BERT~\citep{DBLP:conf/naacl/DevlinCLT19}. The symbols $\text{Q}_{i}$ and $\text{C}_{j}$ denote subword tokens of the query identifier and code, respectively. For simplicity, we do not explicitly show the special delimiter tokens such as \codeID{[CLS]}. The input sequence is fed to the pre-trained encoder. 
The span prediction layer consists of a token classifier that performs a four-way classification over the labels $\{B, I, O, F\}$.
It is applied to the encoding of every code token in the last layer of the encoder. 
For negative examples, all tokens are to be classified as $O$.

\paragraph{Two-step procedure of relevance classification and span prediction.}
We found that in many cases, the entire file contents do not fit in the input to the model. However, not all code is relevant for answering a given query. 
As discussed in Section~\ref{sec:dataset-preparation}, we identify the relevant code blocks programmatically using the CodeQL results during data preparation.
We use this information to devise a two-step procedure (see Figure~\ref{fig:two-step}) to deal with the problem of scaling to large-size code: 
\begin{enumerate}
\vspace*{-3mm}
    \item[\emph{Step 1}:] We first apply a \emph{relevance classifier} to every block in the given code and select code blocks that are likely to be relevant for answering a given query.
    \vspace*{-1.5mm}
    \item[\emph{Step 2}:] We then apply the span prediction model (Figure~\ref{fig:model-architecture}) to the set of selected code blocks to predict answer and supporting-fact spans.
    \vspace*{-1mm}
\end{enumerate}

\emph{Training}:
Let $F$ be a file and $R$ be the set of code blocks in $F$ that are relevant for a query $Q$. Other blocks in $F$ are irrelevant. We train a classifier that given $Q$ and a code block $b$ predicts whether $b$ is relevant or not. We fine-tune a CuBERT checkpoint as the relevance classifier.
%
Instead of training the span prediction model on the entire contents of a file $F$, we train it on code blocks relevant for $Q$ within $F$. The code blocks identified as relevant during data preparation are used for training. We fine-tune the models by minimizing the cross-entropy loss. 

\emph{Inference}:
At inference time, given a query $Q$ and a file comprising code blocks $\{b_1, \ldots, b_n\}$, we generate a set of $n$ examples by concatenating $Q$ and the contents of each of $b_i$. The relevance classifier is applied on each of these examples and all blocks classified as relevant are selected. The selected blocks and the query are passed to the span prediction model as shown in Figure~\ref{fig:two-step}.



\subsection{Evaluation Metrics}


We measure the performance of the model in terms of \emph{exact match}. A exact match occurs when the set of predicted answer spans is same as the set of ground-truth answer spans. When supporting facts are predicted, the exact match also requires that the set of predicted supporting-fact spans is same as the set of ground-truth supporting-fact spans. For a relevance classification model, we measure the usual classification metrics: accuracy, precision, and recall.

\section{Experimental Results}
\label{sec:results}

\begin{table}[t]
\centering
\begin{subtable}[@{}c@{}]{0.52\textwidth}
\centering
\scalebox{0.9}{
    \begin{tabular}{@{}r|r|r|r|r@{}}
    \hline
    & \multicolumn{2}{c}{\begin{tabular}[c]{@{}l@{}}Zero-shot prompting\end{tabular}} 
    &  \multicolumn{2}{|c}{\begin{tabular}[c]{@{}l@{}}Few-shot prompting\\ with BM25 retrieval\end{tabular} } \\ \cline{2-5}
    
    Pass@$k$ & Positive & Negative & Positive & Negative\\ \hline
    
     1 & 9.82      & 12.83 & \textbf{16.45} & \textbf{44.25}   \\ 
     2  & 13.06    & 17.42 & \textbf{21.14}    & \textbf{55.53} \\
     5  & 17.47    & 22.85 & \textbf{27.69}    & \textbf{65.43}\\
     10 & 20.84    & 26.77  & \textbf{32.66}    & \textbf{70.08} \\
     \hline
    \end{tabular}}
    \caption{Zero-shot prompting and few-shot prompting with BM25 retrieval for answer span prediction. }
    \label{tab:gpt-exp-a}
\end{subtable}
\hfill
\begin{subtable}[@{}c@{}]{0.42\textwidth}
\centering
\scalebox{0.9}{
    \begin{tabular}{@{}r|c@{}}
    \hline
    & \multicolumn{1}{c}{\begin{tabular}[c]{@{}l@{}}Few-shot prompting\\ with supporting facts\end{tabular} } \\ \cline{2-2}
    
    Pass@$k$ & Positive \\ \hline
    
     1 &  21.88   \\ 
     2  & 28.06  \\
     5  & 34.94\\
     10 & 39.08  \\
     \hline
    \end{tabular}}
    \caption{Few-shot prompting with supporting facts for answer and supporting-fact span prediction.}
    \label{tab:gpt-exp-b}
\end{subtable}
\vspace{-2mm}
\caption{Percentage exact match achieved by GPT3.5-Turbo on the sampled test data.} 
\end{table}


\subsection{Evaluation of the LLM with Zero-shot and Few-shot Prompting}
\label{sec:llm-eval}

\paragraph{Sampled test data.}
Due to a limited inference budget, we evaluate the LLM (GPT3.5-Turbo) on a sample of the test split. Considering the available prompt size of 4096 tokens in the used LLM, we sampled files that can fit into the input along with the examples of few-shot prompts, i.e., files having less than 2000 tokens are considered. For each of the 52 queries, we select a maximum of 20 test files with 10 each from positive and negative examples. We refer to this as the \emph{sampled test data}.

\paragraph{Results on the sampled test data.}
We experiment on the sampled test data with various prompts and obtain 10 generations at temperature of 0.8 per inference. We use the $pass@k$ measure from~\citep{chen_evaluating_2021_Codex} for $k$ draws from $n$ generations, for $k \in \{1,2,5,10\}$ and $n=10$.

Table~\ref{tab:gpt-exp-a} shows the results of zero-shot prompting and few-shot prompting with BM25 retrieval for answer span prediction. 
In zero-shot prompting, the LLM gets only 9.82\% and 12.83\% exact match on positive and negative examples respectively with pass@1.
For $k=10$, these increase to 20.84\% and 26.77\% respectively.
The few-shot prompting shows improvement over zero-shot prompting at all values of $k$. The improvement on negative examples is particularly significant. We believe that this is because both a positive and a negative example are provided in the prompt. The negative example has a plausible but incorrect candidate answer (see Section~\ref{sec:dataset-preparation}). The  difference in the two examples helps the LLM detect the negative examples more accurately.

Table~\ref{tab:gpt-exp-b} shows the results of few-shot prompting with supporting facts on answer and supporting-fact span prediction. As discussed in Section~\ref{sec:prompting}, this setting is applicable only to positive examples. We see that the LLM achieves exact match of 21.88\%--39.08\% for different values of $k$.
Note that for the experiment in Table~\ref{tab:gpt-exp-a}, the model is required to distinguish between positive and negative examples, which is not the case in this setting. The additional annotation of supporting facts in the examples in the prompt seems to help the model in predicting both answers and supporting facts.

\paragraph{Observations.}
With zero-shot prompting, the LLM was able to identify correct spans in positive examples for simple queries, e.g., 80\% exact match for the query ``Flask app is run in debug mode'', but achieved no exact match on complex queries like ``Inconsistent equality and hashing''. 
It faces similar problems with the negative examples.
Some of these failure cases are fixed with few-shot prompting where explicit spans of positive/negative examples in the prompt provide additional information about the intent and differences between positive/negative examples. For many queries including ``Inconsistent equality and hashing'', few-shot prompts having examples with supporting facts are able to generate correct answer spans along with the correct supporting facts. As general observations, for both single-hop and multi-hop queries, we see shorter and more accurate code generation with few-shot prompts compared to zero-shot prompts.

\subsection{Evaluation of the Fine-tuned Contextual Embedding Models}
\label{subsec:eval-twostep}

\begin{table}[t]
\centering
\begin{subtable}[@{}c@{}]{0.4\textwidth}
\centering
\scalebox{0.9}{
\begin{tabular}{@{}l@{\;}|@{\;}r@{\;}|@{\;}r@{}}
\hline
Variants & Positive & Negative\\
\hline
Two-step(20, 20)   & 3.74 & 95.54 \\
Two-step(all, 20)   &  7.81 & 97.87 \\ 
Two-step(20, all)   & 33.41 & 96.23 \\  
Two-step(all, all) & \textbf{52.61}  & \textbf{96.73} \\  \hline\hline
Prefix      & 36.60   & 93.80   \\ 
Sliding window  & 51.91  & 85.75  \\ \hline
\end{tabular}}
\caption{Answer and supporting-fact span prediction on the complete test data.}
\label{tab:cubert-a}
\end{subtable}
\hfill
\begin{subtable}[@{}c@{}]{0.58\textwidth}
\centering
\scalebox{0.9}{
    \begin{tabular}{@{}l@{\;}|@{}r@{\;}|@{}r@{\;}||@{}r@{}}
    \hline
    & \multicolumn{2}{c||}{\begin{tabular}{@{}c@{}}Answer span\\ prediction\end{tabular}} & 
    \begin{tabular}{@{}c@{}}Answer \& supporting-\\fact span prediction\end{tabular}\\ \cline{2-4}
     Variants & Positive & Negative & Positive\\ \hline
    \begin{tabular}[c]{@{}l@{}}Two-step(20, 20)\end{tabular}       &  9.42        &    92.13 & 8.42     \\ \hline
    \begin{tabular}[c]{@{}l@{}}Two-step(all, 20)\end{tabular}       &  15.03        &    94.49 &   13.27     \\ \hline
    \begin{tabular}[c]{@{}l@{}}Two-step(20, all)\end{tabular}       &  32.87        &    96.26  &   30.66    \\ \hline
    \begin{tabular}[c]{@{}l@{}}Two-step(all, all)\end{tabular}      &  51.90       &    95.67  &    49.30    \\  \hline
    \end{tabular}}
    \caption{Results on the sampled test data from Section~\ref{sec:llm-eval}.}
    \label{tab:cubert-b}
\end{subtable}
\vspace*{-2mm}
\caption{Percentage exact match achieved by the models fine-tuned from CuBERT.}
\label{tab:llm-exp-twostep}
\end{table}

\paragraph{Training setup.}
We fine-tune the relevance classification and span prediction models from the pre-trained CuBERT checkpoints for 512 and 1028 token lengths respectively. Each of them is trained jointly on all 52 queries. We train two variants each of these models: 1)~one on \emph{all} files in the training split and 2)~another on \emph{10 positive and 10 negative files per query} as a representative of the practical setting in which only a few labeled examples are available. 
We denote the resultant two-step procedure (classification followed by span prediction) by \emph{two-step$(x, y)$} indicating that the relevance classifier is trained with $x$ files and the span predictor is trained with $y$ files from the training data, for $x, y \in$ \{20, all\}. We provide the full details of the training setup in the supplementary material.

\paragraph{Results on the complete test data.}
As these models are run locally, we can evaluate them on the complete test data (unlike the LLM).
Table~\ref{tab:cubert-a} gives results of the two-step procedure on the complete test data.
The two-step(all, all) setup which uses all the training data for both the relevance classification and span prediction performs the best, getting 52.61\% and 96.73\% exact match on positive and negative examples. However, it relies on existence of a large set of labeled examples for training, which may not be available in practice. The most practical setting, two-step(20,20), is able to get exact match on only 3.74\% positive examples. Among the $\{B, I, O, F\}$ labels, the label \textbf{O}utside is very frequent compared to the other labels and hence, the token classifier is biased towards predicting it and that explains why the exact match is high for the negative examples in all settings.

The relevance classifier trained with 20 files achieves accuracy, precision, and recall scores of 91.37, 79.72, and 89.61, respectively. Training it with all files increases the scores to 96.38, 95.73, and 90.10 respectively. We evaluated two simple substitutes to relevance classification in the two-step procedure. We considered a \emph{prefix} setup in which the maximum file prefix that can fit the input is selected. Another setup is a \emph{sliding window} setup in which a file is split by the input size of the model into different chunks forming independent examples and the results are aggregated across the chunks. Table~\ref{tab:cubert-a} shows the results obtained by the span prediction model, trained on \emph{all} data, in conjunction with prefix/sliding window. We see that two-step(all,all) performs better than them.

\paragraph{Results on the sampled test data.}
Table~\ref{tab:cubert-b} gives results of the two-step procedure on the sampled test data from Section~\ref{sec:llm-eval}.
We see that two-step(20, 20) has comparable performance to the LLM in pass@1 in zero-shot prompting on answer-span prediction over positive examples (Table~\ref{tab:gpt-exp-a}).
It underperforms the LLM for higher values of $k$ and in few-shot prompting, including for predicting both answer and supporting-fact spans (Table~\ref{tab:gpt-exp-b}).
Increasing the training budget to \emph{all} examples improves the performance of the fine-tuned models.
As discussed earlier, the high performance on negative examples is an artifact of the skew in the token labels towards the \textbf{O}utside label.

\paragraph{Observations.}
For some queries like ``Imprecise assert'' a single file may contain multiple candidate answer spans, e.g., multiple assert statements. With limited training, the relevance classifier had low recall, missing out on some of the relevant candidates. 
Training with more data allows the relevance classifier to avoid considering irrelevant code blocks as relevant, which can be observed in the significant increase in precision score. For single-hop queries, most of the code blocks in a file would be irrelevant. Training with more data resulted in a significant boost ($\geq$ 10\%) in accuracy score for 15 single-hop queries.
For some queries such as ``Module is imported with `import' and `import from''', there is less ambiguity in relevant versus irrelevant blocks and those queries did not benefit much from larger training data.

The span prediction model trained on limited data achieves some success only on a few queries where the answer spans follow specific syntactic patterns, e.g., ``Deprecated slice method'' whose answer spans contain one of \codeID{\_\_getslice\_\_}, \codeID{\_\_setslice\_\_} or \codeID{\_\_delslice\_\_}. On these queries, training on larger data does not improve the model performance much. 
In general, the span prediction works better on single-hop queries than multi-hop queries, even when trained on all data.

\section{Discussion}

\emph{Compute}:
All experiments with fine-tuned models were performed on a 64 bit Debian system with an NVIDIA Tesla A100 GPU having 40GB GPU memory and 85GB RAM.
For evaluating GPT3.5-Turbo, we used the Azure OpenAI service.
%
\emph{Limitations}:
Our dataset consists of 52 queries spanning those many distinct program analysis tasks. There are other queries in the CodeQL suites that can be added in future. We create a dataset over Python code. We are releasing our data preparation code that can be extended to support more queries and more programming languages. Our evaluation is limited to two models, but they are representative of the popular classes of encoder-only and decoder-only pre-trained models. We consider file-level context but there is scope to increase it to include entire code repositories.
%
\emph{Societal impact}:
Neural models are increasingly used as coding assistants. As the assistants evolve into more autonomous agents, it is important to evaluate the depth and accuracy of semantic understanding of the neural models. This can help increase trustworthiness of these models and benefit the society by producing more reliable software.

\section{Conclusions and Future Work}
\label{sec:conclusions}

We presented the \dataset dataset to test the ability of neural models to understand code semantics on the proposed problem of answering semantic queries over code. It requires a model to perform single- or multi-hop reasoning, understand structure and semantics of code, distinguish between positive and negative examples, and accurately identify answer and supporting-fact spans. Our evaluation shows that \dataset is challenging for the best-in-class generative and embedding approaches under different prompting or fine-tuning settings.
We are considering extensions to our dataset to include more semantic queries and more programming languages.


\bibliography{references}
\bibliographystyle{neurips_2022}

\newpage

\appendix

\setcounter{page}{1}
\resetlinenumber

\newcommand{\mytoptitlebar}{
  \hrule height 4pt
  \vskip 0.25in
  \vskip -\parskip%
}
\newcommand{\mybottomtitlebar}{
  \vskip 0.29in
  \vskip -\parskip
  \hrule height 1pt
  \vskip 0.09in%
}

\begin{table}[t]
\mytoptitlebar
\centering
{\LARGE\bf Supplementary Material on \dataset\par}
\mybottomtitlebar
\end{table}

\section{Additional Details}
\subsection{Comparison to Existing Datasets}
Existing datasets for question-answering in the context of programming languages target comparatively simpler tasks of predicting binary yes/no answers to a question or range over a localized context (e.g., a source-code method). In contrast, in \dataset, a source-code file is annotated with the required spans for a code analysis query about semantic aspects of code. Such a dataset can be used to experiment with various methodologies in an extractive question-answering setting with file-level code context. We tabulate a brief comparison of existing datasets considered for question-answering tasks on source code in Table~\ref{tab:code-datasets}.

\begin{table}[!h]
\footnotesize
\centering
    \begin{tabularx}{\textwidth}{p{0.15\textwidth}|p{0.16\textwidth}|p{0.25\textwidth}|p{0.2\textwidth}|p{0.15\textwidth}}
    \hline
    \textbf{Dataset} & \textbf{Size (Language)} & \textbf{Task} & \textbf{Evaluation Criteria} &
      \textbf{Code\newline Context} \\ 
    \hline
    \hline
    
    CoSQA~\citep{DBLP:conf/acl/HuangTSG0J0D20} &   20,604 (Python) & To check relevance between a web query and a method. & MRR & Method \\ \hline
    CodeQA~\citep{DBLP:conf/emnlp/Liu021} & 119,778 (Java) \newline 70,085 (Python)&   To generate free-form answers for template-based questions curated from comments. & BLEU, \newline ROUGE-L,\newline METEOR,\newline Exact Match,\newline F1  & Method \\ \hline
    Bansal et. al.~\citep{DBLP:conf/wcre/BansalEWM21} & $\approx$10880K (Java) &
      To answer template-based basic questions on method characteristics & User study &
      Method \\ \hline
    CS1QA~\citep{lee2022cs1qa} & 9,237 (Python) &  To classify the question into pre-defined types, identify relevant source code lines and retrieve related questions & Accuracy,\newline F1,\newline Exact Match (line-level)\newline  & Method \\ \hline
    \textbf{\dataset} (this work) &   133,456 (Python) See Tables~\ref{tab:querywise-multihop}--\ref{tab:querywise-singlehop} for details. & To extract answer spans from a given code context in response to a code analysis query, and provide reasoning with supporting-fact spans. & Exact Match &
      File \\ \hline 
      
    \end{tabularx}
\caption{Comparison to existing datasets on question-answering over source code.
}
\label{tab:code-datasets}
\end{table}

\subsection{Query-wise Dataset Statistics}
\label{sec:querywise-statistics}
We report the query-wise statistics for multi-hop and single-hop queries, aggregated across all splits, in Table~\ref{tab:querywise-multihop} and Table~\ref{tab:querywise-singlehop} respectively. 
We report the statistics for \emph{All Examples}, \emph{Positive} examples, and \emph{Negative} examples. \emph{Count} gives the number of examples. A single file may be part of examples of multiple queries. Each example in Table~\ref{tab:querywise-multihop} and Table~\ref{tab:querywise-singlehop} corresponds to a query and file pair, whereas Table~\ref{tab:file-data-stat} tabulates the number of \emph{unique} files in different splits of the dataset. We sort all the tables from here on by the descending order of the count of all examples. Under all examples, we give the average length of the input sequences in terms of sub-tokens. Here, the sub-tokenization is performed using the CuBERT vocabulary. For positive examples, we report the average number of answer (abbreviated as \emph{Ans.}) spans and supporting fact (abbreviated as \emph{SF}) spans. Note that the number of answer or supporting fact spans is zero for negative examples and are hence omitted. We highlight the minimum and maximum values per column in bold face.

\newcolumntype{R}[1]{>{\raggedleft\let\newline\\\arraybackslash\hspace{0pt}}m{#1}}
\newcolumntype{L}[1]{>{\raggedright\let\newline\\\arraybackslash\hspace{0pt}}m{#1}}

{\small
\begin{longtable}[c]{|@{}R{0.8cm}|p{4cm}|@{}R{1.2cm}|R{1.2cm}|@{}R{1cm}|@{}R{1cm}|@{}R{1cm}|@{}R{1.3cm}|}
\caption{Query-wise statistics for the multi-hop queries.}
\label{tab:querywise-multihop}\\
\hline
 \textbf{Index} & \multirow{1}{*}{\textbf{Query Name}} & \multicolumn{2}{c|}{\textbf{All Examples}} & \multicolumn{3}{c|}{\textbf{Positive}} & \textbf{Negative} \\
\cline{3-8}
& & Count & Avg. Length & Count & Avg. Ans. Spans & Avg. SF Spans  & Count \\
\hline
\endfirsthead
\hline
 \textbf{Index} & \multirow{1}{*}{\textbf{Query Name}} & \multicolumn{2}{c|}{\textbf{All Examples}} & \multicolumn{3}{c|}{\textbf{Positive}} & \textbf{Negative} \\
\cline{3-8}
 & & Count & Avg. Length & Count & Avg. Ans. Spans & Avg. SF Spans  & Count \\
\hline
\endhead
\hline
\endfoot
\cline{1-8}
Q1 & \unusedimport & \textbf{48,555} & 3037.87 & \textbf{19,178} & 2.1 & \textbf{0} & \textbf{29,377}\\
Q2 & \missinginitduringinit & 1,115 & 6860.32 & 353 & 2.18 & 3.06 & 762\\
Q3 & \useofnonereturn & 919 & 6514.05 & 348 & 1.67 & 1.02 & 571\\
Q4 & \wrongargincall & 700 & 8266.05 & 272 & 1.61 & 1.12 & 428\\
Q5 & \eqnotoverriden & 547 & 8429.84 & 500 & 1.56 & \textbf{5.61} & \textbf{47}\\
Q6 & \comparisonusingis & 453 & 10136.81 & 151 & 2.05 & \textbf{0} & 302\\
Q7 & \noncallablecalled & 375 & 9362.16 & 118 & 2.23 & 1.84 & 257\\
Q8 & \signmismatch & 374 & 11245.87 & 127 & \textbf{2.32} & 1.32 & 247\\
Q9 & \initcalloverriden & 371 & \textbf{11335.33} & 176 & 1.31 & 4.44 & 195\\
Q10 & \iterreturnnoniter & 266 & 9196.37 & 165 & 1.27 & 1.36 & 101\\
Q11 & \conflictingattr & 255 & 8920.37 & 96 & 1.9 & 3.07 & 159\\
Q12 & \flaskdebug & 242 & \textbf{1134.98} & 123 & \textbf{1.0} & \textbf{0} & 119\\
Q13 & \unequalhash & 195 & 9964.23 & 100 & 1.21 & 1.21 & 95\\
Q14 & \wrongarginclassinst & 188 & 7608.82 & \textbf{79} & 1.46 & 0.96 & 109\\
Q15 & \incompleteordering & \textbf{153} & 9628.29 & 80 & 1.09 & 1.43 & 73\\

\hline
 \multicolumn{2}{|c|}{Aggregate} & 54,708 & 3617.26 & 21,866 & 2.04  & 0.30 & 32,842      
\end{longtable}
}

Table~\ref{tab:querywise-singlehop} gives the query-wise statistics for single-hop queries aggregated across all splits. The column headings have the same meaning as those of Table~\ref{tab:querywise-multihop}. We highlight the minimum and maximum values per column in bold face.

{\small
\begin{longtable}[c]{|@{}R{0.8cm}|p{4cm}|@{}R{1.2cm}|R{1.2cm}|@{}R{1cm}|@{}R{1cm}|@{}R{1cm}|@{}R{1.3cm}|}
\caption{Query-wise statistics for the single-hop queries.}
\label{tab:querywise-singlehop}\\
\hline
 \textbf{Index} & \multirow{1}{*}{\textbf{Query Name}} & \multicolumn{2}{c|}{\textbf{All Examples}} & \multicolumn{3}{c|}{\textbf{Positive}} & \textbf{Negative} \\
\cline{3-8}
& & Count & Avg. Length & Count & Avg. Ans. Spans & Avg. SF Spans  & Count \\
\hline
\endfirsthead
\hline
 \textbf{Index} & \multirow{1}{*}{\textbf{Query Name}} & \multicolumn{2}{c|}{\textbf{All Examples}} & \multicolumn{3}{c|}{\textbf{Positive}} & \textbf{Negative} \\
\cline{3-8}
 & & Count & Avg. Length & Count & Avg. Ans. Spans & Avg. SF Spans  & Count \\
\hline
\endhead
\hline
\endfoot
\cline{1-8}
Q16 & \unusedlocalvar & \textbf{22,711} & 5399.66 & \textbf{8,123} & 2.53 & 0 & \textbf{14,588}\\
Q17 & \baseexception & 14,893 & 5081.62 & 5,909 & 2.23 & \textbf{0} & 8,984\\
Q18 & \vardefmultiple & 8,548 & 7147.93 & 2,596 & 2.58 & 1.94 & 5,952\\
Q19 & \impreciseassert & 6,699 & \textbf{4089.02} & 2,192 & 5.67 & \textbf{0} & 4,507\\
Q20 & \unreachablecode & 4,146 & 8025.58 & 1,726 & 1.46 & \textbf{0} & 2,420\\
Q21 & \testeqnone & 4,045 & 8100.94 & 1,408 & 2.27 & \textbf{0} & 2,637\\
Q22 & \firstparaself & 2,357 & 8031.02 & 444 & 4.6 & \textbf{0} & 1,913\\
Q23 & \doubletypeimport  & 1,918 & 5057.49 & 912 & 1.11 & \textbf{0} & 1,006\\
Q24 & \unnecessarypass & 1,812 & 7902.1 & 757 & 1.86 & \textbf{0} & 1,055\\
Q25 & \doubleimport & 953 & 5384.63 & 391 & 1.45 & 1.13 & 562\\
Q26 & \constcompare & 839 & 10276 & 61 & \textbf{13.72} & \textbf{0} & 778\\
Q27 & \implicitstringconcat & 787 & 8942.5 & 237 & 2.35 & \textbf{0} & 550\\
Q28 & \unusedloopvar & 750 & 9927.05 & 317 & 1.36 & \textbf{0} & 433\\
Q29 & \duplicatedictkey & 675 & 8655.74 & 131 & 4.56 & \textbf{4.37} & 544\\
Q30 & \unnecessaryelse & 606 & 9305.47 & 278 & 1.24 & \textbf{0} & 328\\
Q31 & \redundantassign & 566 & 7991.59 & 231 & 1.46 & \textbf{0} & 335\\
Q32 & \firstargsuper & 560 & 5322.95 & 236 & 1.4 & \textbf{0} & 324\\
Q33 & \importdeprecated & 500 & 5889.08 & 228 & 1.19 & \textbf{0} & 272\\
Q34 & \samevarinnestedloop & 496 & 9117.3 & 222 & 1.26 & 1.14 & 274\\
Q35 & \redundantcomp & 425 & 10775.71 & 153 & 1.81 & 1.6 & 272\\
Q36 & \assertsideeffect & 408 & 6800.28 & 109 & 3.12 & \textbf{0} & 299\\
Q37 & \importstar & 397 & 5441.52 & 197 & \textbf{1.02} & \textbf{0} & 200\\
Q38 & \constincondition & 377 & 9761.03 & 118 & 2.19 & \textbf{0} & 259\\
Q39 & \compidentical & 358 & 9861.62 & 108 & 2.32 & \textbf{0} & 250\\
Q40 & \illegalraise & 342 & 7482.82 & 141 & 1.43 & \textbf{0} & 201\\
Q41 & \notimplemented & 340 & 6763.09 & 124 & 1.93 & \textbf{0} & 216\\
Q42 & \unnecessarydelete & 309 & 8875.78 & 146 & 1.36 & 1.36 & 163\\
Q43 & \deprecatedslice & 285 & 10171.74 & 86 & 2.6 & \textbf{0} & 199\\
Q44 & \insecuretemp & 249 & 6488.8 & 107 & 1.41 & \textbf{0} & 142\\
Q45 & \moddefaultpara & 230 & 9112.3 & 88 & 1.61 & 1.23 & 142\\
Q46 & \usewith & 204 & 6525.19 & 91 & 1.26 & 0.02 & 113\\
Q47 & \useglobalmodule & 182 & 6614.24 & 72 & 1.69 & \textbf{0} & 110\\
Q48 & \nonstandardexc & 167 & 9277.62 & 65 & 1.58 & 0.14 & 102\\
Q49 & \moddictlocals & 165 & 7130.5 & 65 & 1.51 & \textbf{0} & 100\\
Q50 & \specialmethodsign & 164 & \textbf{11703.91} & 56 & 1.98 & 1.48 & 108\\
Q51 & \incompleteurlsanity & 154 & 5805.74 & 62 & 1.61 & \textbf{0} & 92\\
Q52 & \unguardednext & \textbf{131} & 7526.71 & \textbf{54} & 1.52 & \textbf{0} & \textbf{77}\\
\hline
 \multicolumn{2}{|c|}{Aggregate} & 78,748  & 6228.49  & 28,241 & 2.51  & 0.25  & 50,507
\end{longtable}
}

\subsection{Statistics of Syntactic Patterns of Spans}
\label{sec:span-stats}
In our dataset, the answer and supporting-fact spans cover various types of programming language constructs. Hence, in \cref{tab:span-type-stat}, we tabulate the number of spans in terms of syntactic patterns of Python constructs in decreasing order of their frequency in the combined data of all three splits. To find the pattern of a span, we have used \codeID{tree-sitter}~\citep{treeSitter} to get the closest ancestor node which encloses the tokens appearing in the span. Two special entries in the table are block and module. A \emph{block} node can represent any block of code, i.e., a block of code, a function, a class. Sometimes the closest ancestor node is the root node of the source code, for those cases \emph{module} node is used as a representative node.

{\small
\begin{longtable}[c]{|p{2.8cm}|r|p{2.9cm}|r|p{2.8cm}|r|}
\caption{Statistics of syntactic patterns of spans.}
\label{tab:span-type-stat}\\
\hline
\textbf{Syntactic Pattern} & \textbf{Count} & \textbf{Syntactic Pattern} & \textbf{Count} & \textbf{Syntactic Pattern} & \textbf{Count}   \\
\hline
\endfirsthead
\hline
Span Type & Count & Span Type & Count & Span Type & Count \\
\hline
\endhead
\hline
\endfoot
\cline{1-6}
import statement & 43,013 & raise statement & 375 & module & 56 \\
assignment & 32,422 & function parameters & 373 & dictionary keys & 47 \\
call & 15,978 & assert statement & 368 & break statement & 43 \\
except clause & 13,269 & delete statement & 358 & while statement & 43\\
function definition  & 8,937  & if statement & 243 & argument list & 34 \\
non-\-boolean binary operator & 5,319 & sequence expressions & 192 & with statement & 26 \\
class attributes & 2,844 & identifier & 186  & parenthesized expression & 14 \\
class definition & 2,882 & decorator & 138 & boolean operator & 13 \\
block & 2,331  & print statement & 126 & elif clause & 12 \\
pass statement & 1,451  & global statement & 125 & expression list & 12 \\
string literal & 1,279  & list comprehension  & 101 &  lambda & 11\\
for statement & 1,164  & subscript & 72  & conditional expression & 8  \\
concatenated string  & 558   & not operator & 71  & yield & 5  \\
return statement & 395 & try statement & 65  & continue statement & 3 \\ \hline
 \multicolumn{4}{|c }{} &  Aggregate & 134,962 \\ \hline
\end{longtable}
}

\subsection{Prompt Templates}
In this section, we provide various prompts used with the GPT3.5-Turbo model. The templates for zero-shot prompting, few-shot prompting with BM25 retrieval, and few-shot prompting with supporting facts are provided in Figure~\ref{prompt:zero-shot}, Figure~\ref{prompt:few-shot-bm25}, and Figure~\ref{prompt:few-shot-sf}, respectively. Few-shot prompting with supporting facts uses two prompt sub-templates given in Figure~\ref{prompt:few-shot-sf-ex-a} and Figure~\ref{prompt:few-shot-sf-ex-b} to add examples with/without supporting facts.
\begin{table}[h]
    \centering
\begin{lstlisting}[
basicstyle={\scriptsize\ttfamily},
identifierstyle={\color{black}},
numbers=none,
language=no_lang,
breaklines=true,
breakindent=0em,
frame=single,
% caption={My Caption},captionpos=b
]
You are an expert software developer. Please help identify the results of evaluating the CodeQL query titled "{{ query_name }}" on a code snippet. The results should be given as code spans or fragments (if any) from the code snippet. The description of the CodeQL query "{{ query_name }}" is - {{ description }}

If there are spans that match the query description, print them out one per line. If no spans matching the query description are present, say N/A.

Code snippet
```python
{{ input_code }}
```

Code span(s)
```python
\end{lstlisting}
\captionof{figure}{Zero-shot prompt template.}
\label{prompt:zero-shot}
\end{table}

\begin{table}[t]
    \centering
\begin{lstlisting}[
basicstyle={\scriptsize\ttfamily},
identifierstyle={\color{black}},
numbers=none,
language=no_lang,
breaklines=true,
breakindent=0em,
frame=single
]
You are an expert software developer. Please help identify the results of evaluating the CodeQL query titled "{{ query_name }}" on a code snippet. The results should be given as code spans or fragments (if any) from the code snippet. The description of the CodeQL query "{{ query_name }}" is - {{ description }}

If there are spans that match the query description, print them out one per line. If no spans matching the query description are present, say N/A.

The following are some examples of code snippets with and without spans matching the query description.
Example code snippet with span(s) matching the query description
```python
{{ positive_context }}
```

Code span(s)
```python
{% for span in positive_spans %}
{{span}}
{% endfor %}
```

Example code snippet with no span(s) matching the query description
```python
{{ negative_context }}
```

Code span(s)
```python
N/A
```

Code snippet
```python
{{ input_code }}
```

Code span(s)
```python
\end{lstlisting}
\captionof{figure}{Few-shot prompt template with BM25 retrieval.}
\label{prompt:few-shot-bm25}
\end{table}

\begin{table}[h]
    \centering 
\begin{lstlisting}[
basicstyle={\scriptsize\ttfamily},
identifierstyle={\color{black}},
numbers=none,
language=no_lang,
breaklines=true,
breakindent=0em,
frame=single
]
You are an expert software developer. Please help identify the results of evaluating the CodeQL query titled "{{ query_name }}" on a code snippet. The results should be given as code spans or fragments (if any) from the code snippet. The description of the CodeQL query "{{ query_name }}" is - {{ description }}

The results should consist of two parts: answer spans and supporting fact spans. If there are spans that match the query description, print them out as answer spans. Supporting fact spans are spans that provide additional evidence about the correctness of the answer spans. Always print one span per line. If no such spans exist, print N/A.

The following are some examples of code snippets with spans matching the query description, along with supporting facts if any.

{ex_a}

{ex_b}

Code snippet
```python
{{ input_code }}
```

Answer span(s)
```python
\end{lstlisting}
\captionof{figure}{Few-shot prompt template with supporting facts.}
\label{prompt:few-shot-sf}
\end{table}

\begin{table}[h]
    \centering
\begin{minipage}[b]{0.45\linewidth}
\begin{lstlisting}[
basicstyle={\scriptsize\ttfamily},
identifierstyle={\color{black}},
numbers=none,
language=no_lang,
breaklines=true,
breakindent=0em,
frame=single
]
Example code snippet with answer span(s) matching the query description with supporting fact span(s)
```python
{{ positive_context }}
```

Answer span(s)
```python
{% for span in positive_spans %}
{{span}}
{% endfor %}
```

Supporting fact span(s)
```python
{% for span in supporting_fact_spans %}
{{span}}
{% endfor %}
```END

\end{lstlisting}
\captionof{figure}{``\codeID{ex\_a}'' sub-template in few-shot prompt with supporting facts.}
\label{prompt:few-shot-sf-ex-a}
\end{minipage}
\hfill
\begin{minipage}[b]{0.45\linewidth}
\begin{lstlisting}[
basicstyle={\scriptsize\ttfamily},
identifierstyle={\color{black}},
numbers=none,
language=no_lang,
breaklines=true,
breakindent=0em,
frame=single
]
Example code snippet with answer span(s) matching the query description but without supporting fact span(s)
```python
{{ positive_context }}
```

Answer span(s)
```python
{% for span in positive_spans %}
{{span}}
{% endfor %}
```

Supporting fact span(s)
```python
N/A
```END

\end{lstlisting}
\captionof{figure}{``\codeID{ex\_b}'' sub-template in few-shot prompt without supporting facts.}
\label{prompt:few-shot-sf-ex-b}
\end{minipage}
\end{table}

    





    

\subsection{Training Setup}
\label{sec:training-setup-appendix}
This section documents the setup used for training the models discussed in Section~\ref{subsec:eval-twostep}. The pre-trained CuBERT encoder model checkpoints are available for input length of 512 and 1024. We use the 1024-length checkpoint for span prediction and the 512-length checkpoint for relevance classification.

For span prediction, the token encodings from the final hidden layer of an encoder are passed through a dropout layer with a dropout probability of 0.1 followed by a classification layer. We initially experimented with up to 10 epochs and learning rates in the order of e-5 and e-6 for these models. We observed that the models reached minimum validation loss with the following configurations and used them. Fine-tuning is performed for 5 epochs for the 512-length models and for 3 epochs for the 1024-length models, with a learning rate of 3e-5. Based on the memory constraints, we used batch sizes of 4 and 16 for sequence lengths 1024 and 512 respectively. All the models are trained by minimizing the cross-entropy loss using the AdamW optimizer~\citep{DBLP:journals/corr/abs-1711-05101} and linear scheduling without any warmup. The best checkpoint is decided based on least validation loss. We used the same hyper-parameters for fine-tuning the CuBERT 1024 span prediction model with a limited number of files (Section~\ref{subsec:eval-twostep}).

For the relevance classification model, we fine-tuned the pre-trained CuBERT model with input length limit of 512. The pooled output is passed through a dropout layer with dropout probability of 0.1 and a 2-layer classifier with a hidden dimension of 2048. We fine-tuned it for 5 epochs with a learning rate of 3e-6 and used weighted crossentropy (with weights 1/2 for irrelevant/relevant class) as the loss function. The best checkpoint is decided based on the least validation loss. We used the same hyper-parameters except for the learning rate (2e-6) for fine-tuning the CuBERT 512 relevance classification model with a limited number of files (Section~\ref{subsec:eval-twostep}).

All experiments are performed on a 64 bit Debian system with an NVIDIA Tesla A100 GPU having 40GB GPU memory and 85GB RAM.

\subsection{Examples of Successful and Unsuccessful Span Predictions}
\label{sec:success-examples}

\begin{table}[h]
\begin{minipage}[b]{0.45\columnwidth}
\begin{lstlisting}[
basicstyle=\small,
frame=single
]
import sys

# Supporting Fact
|\mytikzmark{sf1Start}|class _Registry(dict):|\mytikzmark{sf1End}|
    ...

    def __init__(self):
        dict.__init__(self)
        
    # Answer Span
    |\mytikzmark{spanStart1}|def __hash__(self):|\mytikzmark{spanEnd1}|
        return hash(self.freeze(self))

    def __getitem__(self, key):
        ...

    ...

sys.modules[__name__] = _Registry()    
\end{lstlisting}
\begin{tikzpicture}[remember picture, overlay, x=1mm, y=1mm]
\highlight{spanStart1}{spanEnd1};
\highlightSF{sf1Start}{sf1End};
\end{tikzpicture}
\captionof{figure}{Positive example code labeled with the answer and supporting-fact spans for the ``Inconsistent equality and hashing'' query. }
\label{fig:unequalhash-compressed-1}
\end{minipage}
\hfill
\begin{minipage}[b]{0.45\columnwidth}
\begin{lstlisting}[
basicstyle=\small,
frame=single
]
...

class _AnyLocation:
    ...
    
# Supporting Fact
|\mytikzmark{sf1Start}|class XPathQuery:|\mytikzmark{sf1End}|
    def __init__(self, queryStr):
        ...
        
    # Answer Span
    |\mytikzmark{spanStart1}|def __hash__(self):|\mytikzmark{spanEnd1}|
        return self.queryStr.__hash__()

    def matches(self, elem):
        ...

__internedQueries = {}
...
\end{lstlisting}
\begin{tikzpicture}[remember picture, overlay, x=1mm, y=1mm]
\highlight{spanStart1}{spanEnd1};
\highlightSF{sf1Start}{sf1End};
\end{tikzpicture}
\captionof{figure}{Positive example code labeled with the answer and supporting-fact spans for the ``Inconsistent equality and hashing'' query.}
\label{fig:unequalhash-compressed-2}
\end{minipage}

\label{tab:prmopt_templates}
\end{table}

In this section, we present examples of both successful and unsuccessful predictions of various two-step and LLM prompting setups. \cref{fig:unequalhash-compressed-1}\footnote{Part of \codeID{CenterForOpenScience/scrapi/scrapi/registry.py} file in the ETH Py150 Open dataset} is a positive example of the multi-hop query ``Inconsistent equality and hashing'' where the \codeID{\_\_hash\_\_} method is implemented, but \codeID{\_\_eq\_\_} method is not implemented. Zero-shot prompting fails to generate the answer spans, whereas few-shot prompting with BM25 retrieval and few-shot prompting with supporting facts generate the correct answer span. Among two-step setups, only two-step setups with span prediction models trained with all data, i.e., two-step(20, all) and two-step(all, all), were able to predict the correct spans. \cref{fig:unequalhash-compressed-2}\footnote{Part of \codeID{kuri65536/python-for-android/python-modules/twisted/twisted/words/xish/xpath.py} file in the ETH Py150 Open dataset} is another positive example of the same query, for which all prompting strategies and two-step setups except few-shot prompting with supporting facts failed to predict the answer span.

\begin{minipage}[h]{\columnwidth}
\begin{lstlisting}[
basicstyle=\small,
frame=single
]
...

import subprocess
from helpers import unittest

from luigi.contrib.ssh import RemoteContext

class TestMockedRemoteContext(unittest.TestCase):

    def test_subprocess_delegation(self):
        ...
        |\mytikzmark{spanStart1}|self.assertTrue("ssh" in self.last_test)|\mytikzmark{spanEnd1}|  # Answer Span 1
        |\mytikzmark{spanStart2}|self.assertTrue("-i" in self.last_test)|\mytikzmark{spanEnd2}|  # Answer Span 2
        |\mytikzmark{spanStart3}|self.assertTrue("/some/key.pub" in self.last_test)|\mytikzmark{spanEnd3}|  # Answer Span 3
        |\mytikzmark{spanStart4}|self.assertTrue("luigi@some_host" in self.last_test)|\mytikzmark{spanEnd4}|  # Answer Span 4
        |\mytikzmark{spanStart5}|self.assertTrue("ls" in self.last_test)|\mytikzmark{spanEnd5}|  # Answer Span 5

        subprocess.Popen = orig_Popen

    def test_check_output_fail_connect(self):
        ...
\end{lstlisting}

\begin{tikzpicture}[remember picture, overlay, x=1mm, y=1mm]
\highlight{spanStart1}{spanEnd1};
\highlight{spanStart2}{spanEnd2};
\highlight{spanStart3}{spanEnd3};
\highlight{spanStart4}{spanEnd4};
\highlight{spanStart5}{spanEnd5};
\end{tikzpicture}
\captionof{figure}{Positive example code labeled with the answer spans for the ``Imprecise assert'' query. }
\label{fig:imprecise-assert-compressed-1}
\end{minipage}

\cref{fig:imprecise-assert-compressed-1}\footnote{Part of \codeID{spotify/luigi/test/test\_ssh.py} file in the ETH Py150 Open dataset.} is a positive example of the single-hop query ``Imprecise assert''. For this example, all prompting strategies, i.e., zero-shot prompting, few-shot prompting with BM25 retrieval, and few-shot prompting with supporting facts, were able to generate the correct answer span. Among two-step setups, only two-step setups with span prediction models trained with all data, i.e., two-step(20, all) and two-step(all, all), were able to predict the correct spans.

\begin{table}[h]
\begin{minipage}[h]{0.45\columnwidth}
\begin{lstlisting}[
basicstyle=\small,
frame=single
]
from __future__ import unicode_literals
...

class TypedChoiceFieldTest(SimpleTestCase):
    ...
    def test_typedchoicefield_5(self):
        ...
        self.assertEqual('', f.clean(''))

    def test_typedchoicefield_6(self):
        ...
        self.assertIsNone(f.clean(''))

    def test_typedchoicefield_has_changed(self):
        ...
        self.assertFalse(f.has_changed(None, ''))
        ...
        self.assertTrue(f.has_changed('', 'a'))
        ...

    ...
\end{lstlisting}
\captionof{figure}{Negative example code for the ``Imprecise assert'' query. }
\label{fig:imprecise-assert-neg-compressed-1}
\end{minipage}
\hfill
\begin{minipage}[h]{0.45\columnwidth}
\begin{lstlisting}[
basicstyle=\small,
frame=single
]
from __future__ import unicode_literals
...

class Indent(object):
  ...
  def __init__(self, type, size):
    ...

  def __hash__(self):
    return (self.type, self.size).__hash__()

  def __eq__(self, other):
    return hash(self) == hash(other)

  ...

class GherkinParser(object):
  ...

class GherkinFormatter(object):
  ...
\end{lstlisting}
\captionof{figure}{Negative example code for the ``Inconsistent equality and hashing'' query. }
\label{fig:unequalhash-neg-compressed-1}
\end{minipage}

\label{tab:prmopt_templates}
\end{table}

\cref{fig:imprecise-assert-neg-compressed-1}\footnote{Part of \codeID{ django/django/tests/forms\_tests/field\_tests/test\_typedchoicefield.py} file in the ETH Py150 Open dataset.} is a negative example of the single-hop query ``Imprecise assert''. For this example, zero-shot prompting fails to generate `N/A', whereas few-shot prompting with BM25 retrieval was able to generate the `N/A', denoting the absence of the desired span. Among two-step setups, all setups except two-step(20, 20), were able to predict the absence of spans.

\cref{fig:unequalhash-neg-compressed-1}\footnote{Part of \codeID{waynemoore/sublime-gherkin-formatter/lib/gherkin.py} file in the ETH Py150 Open dataset.} is a negative example of the multi-hop query ``Inconsistent equality and hashing''. For this example, zero-shot prompting and few-shot prompting with BM25 retrieval were not able to generate the required `N/A'. Among two-step setups, all setups except two-step(20, 20), were able to predict the absence of any desired answer spans.

\end{document}